\documentclass[aps,prb,floatfix,reprint,longbibliography]{revtex4-2}

\usepackage[utf8]{inputenc}
\usepackage{amsmath,amsbsy,amssymb,amsfonts}
\usepackage{graphicx, color}
\usepackage{esint}
\usepackage{empheq}
\usepackage{hyperref}
\usepackage{ulem} %Just for editing

\usepackage{appendix} %for appendix/SI numberin

\makeatletter

\usepackage{subfigure}
\usepackage{soul}
\usepackage{xcolor}
\usepackage{soul}
\usepackage{amsthm}
\usepackage{color}
\usepackage{bm}
\usepackage{mathrsfs}
\usepackage{graphics}
\usepackage{lipsum}

\makeatother

\begin{document}

%\preprint{}

%\title{Observed enhanced emission at higher-order exceptional points}
\title{Observed enhanced emission at higher-order exceptional points in RF circuits}

\affiliation{Wave Transport in Complex Systems Lab, Department of Physics, Wesleyan University, Middletown, Connecticut 06459, USA}

\author{Nicolas Wyszkowski$^{1}$, Arunn Suntharalingam$^{1}$, Max Vitek$^{1}$, Arkady Kurnosov$^{1}$, Lucas J. Fern{\'a}ndez-Alc{\'a}zar$^{1,2,\dagger}$,  Tsampikos Kottos$^{1,*}$\\}

\affiliation{
$^{1}$Wave Transport in Complex Systems Lab, Physics Department,\\
Wesleyan University, Middletown, CT-06459, USA\\
$^{2}$Institute for Modeling and Innovative Technology
(CONICET - UNNE) and \\
Natural \& Exact Science Faculty, Northeastern University\\ 
Corrientes, W3404AAS, Argentina\\
 email: $^\dagger$ lfernandez@exa.unne.edu.ar, $^*$ tkottos@wesleyan.edu}

\begin{abstract}
The Purcell effect -- stemming directly from the celebrated Fermi's Golden Rule -- links the enhanced emissivity of an emitter to the 
local density of states (LDoS) of a surrounding cavity. Under typical circumstances the LDoS is assumed to have a Lorentzian lineshape.
Here, we go beyond the traditional Purcell framework by designing RF cavities with non-Lorentzian LDoS caused by
higher-order non-Hermitian exceptional point degeneracies (EPDs) where $N\geq 2$ eigenfrequencies and their associated eigenmodes coalesce. 
We experimentally demonstrate a non-conventional emissivity enhancement (as compared to the isolated resonance regime) that increases with 
the EPD order $N$. The theoretical analysis traces its origin to an $N$-th power Lorentzian LDoS line shape that dominates under judicious 
spatially designed cavity losses. Our results reveal a new route to design cavities that do not rely on ultrahigh $Q$-factor resonators or small 
modal volumes.
\end{abstract}


\maketitle
{\it Introduction --}
The intricate mathematical complexity that underlines the formation of exceptional point degeneracies (EPDs) in the resonant spectrum of 
non-Hermitian systems\cite{Kato95,Ma98,Moiseyev11,Bender18} has resulted in several surprises that challenged our fundamental understanding of wave-matter interactions, offering 
new technological opportunities \cite{EGMKMRC18,FEGG17,ORNY19,MA19}. Chiral mode transfer \cite{DMBKGLMRMR16,XMJH16,ZKCE18,UMM11}, robust wireless power transfer \cite{AYF17,AF20,ZZXQC18}, hypersensitive sensing \cite{W14,HCGSKK17,CWGHEC17,KJLLLZZ22,rodion1,natcomm} (and 
more, see reviews \cite{EGMKMRC18,FEGG17,ORNY19}) are ramifications of EPDs, defined as points in the parameter space of cavities where $N$ eigenfrequencies and 
their corresponding resonant modes coalesce \cite{Kato95,Ma98}. Many of these exotic phenomena have been traced to the perplexing topology of the 
eigenfrequency Riemann surfaces around an EPD where a small perturbation $\epsilon$ leads to a fractional Puiseux series expansion of the 
eigenfrequencies \cite{evamaria}. Much less attention has been directed to the effects of the eigenbasis collapse at an EPD on various physical 
observables. Specifically, it directly affects the line shape of the local density of states (LDoS), which strongly deviates from the Lorentzian form characterizing typical resonances \cite{zin1,PZMHHRSJ17,zin2,RFH21,RFKHH21,ZHOG21,KW20,HBCE22}.

A physical process that is controlled by the LDoS line shape is the emissivity of an emitter embedded inside a cavity. 
An implementation of Fermi's Golden Rule (FGR) for the evaluation of the emission rate $\Gamma$ of a source leads to the conclusion that the emissivity is proportional to the LDoS, which exhibits a Lorentzian profile in the case of 
isolated resonant modes.
In this scenario, at the 
resonant frequency, $\Gamma$ is enhanced (with respect to free space emission) proportionally to the $Q$-factor of the cavity modes; a phenomenon 
known as the Purcell effect \cite{PTP46}. However, this traditional Purcell effect does not consider anomalous cavity conditions 
like the ones associated with resonant EPDs where standard non-degenerate perturbation theory collapses -- as it relies on Taylor expansions 
of differentiable functions \cite{evamaria} while, near EPDs, eigenfrequencies change non-analytically in response to small system perturbations. 
Instead, one needs to use a Jordan-form-based perturbative expansion which results in a non-Lorentzian LDoS line shape that enhances 
emissivity beyond the traditional Purcell effect \cite{PZMHHRSJ17}. Sparse experimental efforts confirmed the anomalous Purcell enhancement near EPDs, but only for second-order ($N=2$) EPDs \cite{arkady,PBGC20, FBPC22}. Of 
particular interest is the implementation of even higher-order EPDs, which could potentially provide a new route for achieving further 
enhancement of the LDoS and narrowing of the emission linewidth (in contrast to traditional Purcell-based methods, which rely on
increasing the $Q$-factor and reducing the mode volume)\cite{WFZ22,HLC25}. Unfortunately, such experiments have so far remained elusive.

\begin{figure}[htbp]
\centering
    \includegraphics[width=\linewidth]{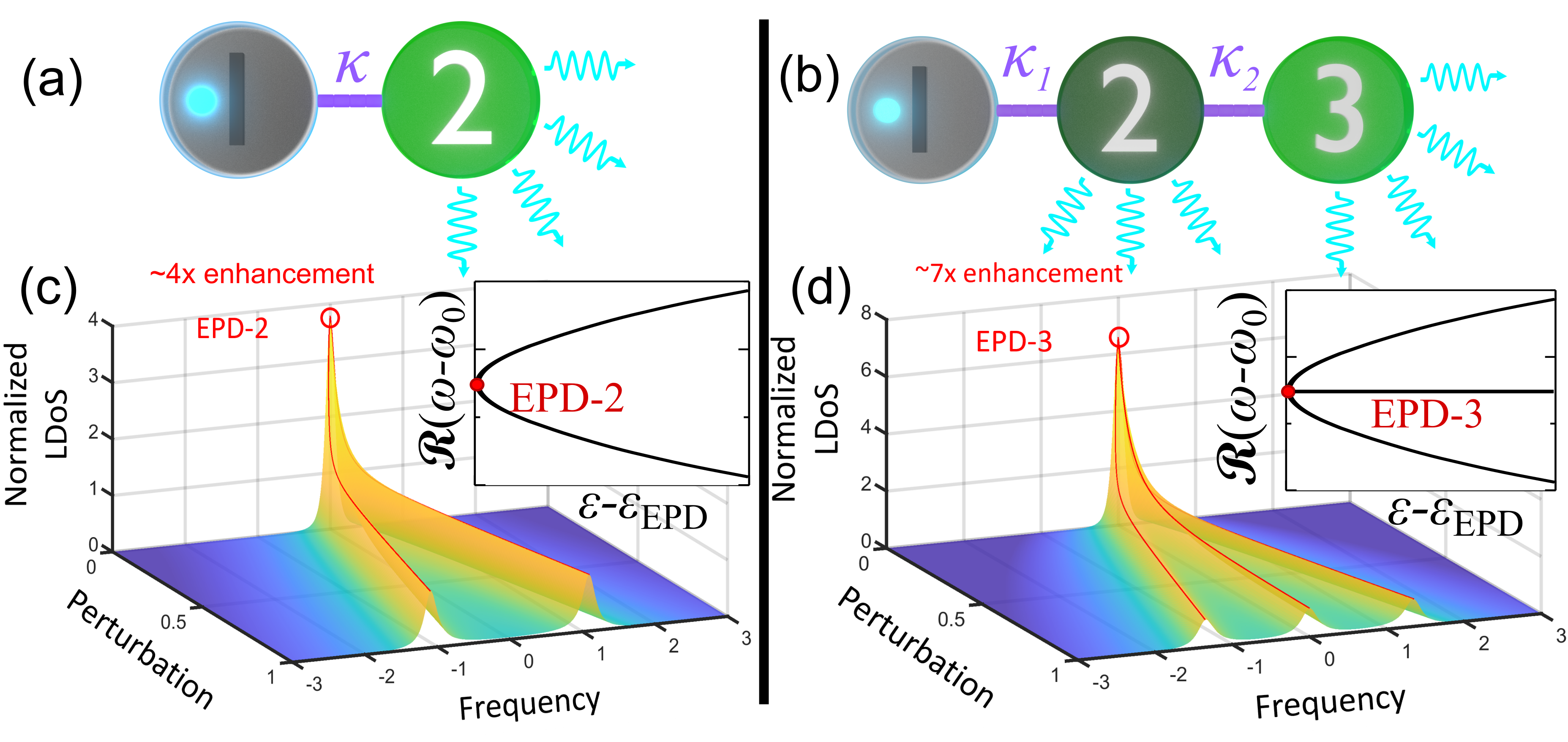}
    \caption{Schematics of a cavity consisting of (a) two and (b) three resonators that are coupled together via variable coupling $\kappa$, 
    forming two and three supermodes. These supermodes coalesce at a critical coupling strength, forming an EPD-$N$ with $N=2,3$, respectively. 
    A monochromatic emitter (light blue circle) is coupled to the cavity via a ``dark'' mode at $n=1$. The emission is measured from the other 
    ``bright'' modes with losses $\Gamma_{n\neq 1}$. Normalized LDoS with its peak-values far from the EPD-$N$ for the (c) two-mode and (d) three-mode cavity versus coupling variations $\kappa=\kappa_{\mathrm{EPD}}+\varepsilon$, displaying an enhancement factor of $4$ and 
    $7$ for $N=2,3$ respectively.}
    \label{fig1}
\end{figure}

Here, we experimentally and theoretically analyze the influence of higher-order resonant EPDs in the emissivity characteristics of a 
source that is coupled to an RF cavity. Specifically, we consider RF cavities consisting of two and three RLC resonators that support 
second ($N=2$)- and third ($N=3$)-order EPDs, respectively (see schematics in Fig. \ref{fig1}), and compare the emissivity for each of 
these cases to their corresponding isolated mode configurations where traditional Purcell physics is applicable. Our experimental results 
show that higher-order $N=3$ EPDs further boost the emissivity beyond the traditional Purcell prediction in accordance with detailed circuit simulations. A coupled 
mode theory traces this $N$-dependent anomalous relative emissivity enhancement to the judicious design of cavity losses that enforces the 
dominant presence of an $N$-th power Lorentzian line shape. Our conclusions are scalable to other frequencies, ranging from elastic 
and acoustic waves to microwave and optical waves. 

%%%%%%%%%%%%%%%%%%

\textit{Coupled Mode Theory Modeling.-} We consider a system with $N=2,3$ resonators (see Fig. \ref{fig1}(a,b)), each supporting one 
resonant mode, which is coupled to an emitting source. The field dynamics is described by a temporal coupled mode theory (CMT) \cite{H84}
\begin{equation}
    \frac{\text{d} \Psi(t)}{\text{d}t}+i H\Psi(t) = v_e S^+(t);\quad  S^+(t)=(S^+_1(t),0\cdots,0)^T 
    \label{eq:CMT}
\end{equation}    
where $\Psi=(\psi_1,\cdots,\psi_N)^T$, with $\psi_n$ being the field amplitude at the $n=1,\cdots,N$-th resonator. $H$ 
is the Hamiltonian that dictates the field dynamics inside the cavity. Its off-diagonal elements $H_{nm}=\kappa_n \delta_{n,m
\pm 1}$ model the coupling between the resonators, and the diagonal elements $H_{nn}=\omega_0 -i\Gamma_n$ model the resonant 
frequencies, $\omega_0$, and the decay rates $\Gamma_{n}=\gamma_n+(WW^T)_{nn}$ due to internal dissipation $\gamma_n$ and 
radiation losses described by the system-environment coupling matrix $W$. Furthermore, we assume that the resonator $n=1$ is a ``dark'' mode, with $\Gamma_1=0$, which 
is coupled directly with coupling strength $v_e\ll \kappa$ (weak source-coupling limit) to an emitting monochromatic source 
of frequency $\omega$ and emitting power $P_e=v_e\cdot |S^+_1|^2$. We want to evaluate the total emitted power from the ``bright'' 
resonators $P_{\rm total}=\sum_{n\neq 1}P_n$ where the emitted power from the $n-$th resonator is $P_n=2\Gamma_n |\psi_n|^2$.

\textit{Fermi's Golden Rule, LDoS and Emitted power -} The FGR determines the rate $\Gamma_{\rm FGR}$ at which the emitter's 
signal enters the cavity via the dark mode, 
\begin{equation}
\label{FGR}
    \Gamma_{\rm FGR}= 2\pi v_e^2 \cdot \xi_1(\omega),\quad      \xi_{1}(\omega)=-\Im[G_{11}(\omega)]/\pi, 
\end{equation}
where $\xi_1(\omega)$ is the LDoS at the dark mode and $G_{mn}$ refers to the $(m,n)$-element of the Green's function  $G(\omega)\equiv(\omega\mathcal{I}-H)^{-1}$. 

The radiation entering the structure from the dark mode $n=1$, $P_{1,{\rm in}}=\Gamma_{\rm FGR}\cdot |S_1^+|^2$, 
will be emitted through the remaining ``bright'' modes 
$n=2,\cdots, N$. Using the optical theorem \cite{PM01}, we can express the diagonal elements of the Green's function that 
appear in Eq. (\ref{FGR}) to the off-diagonal elements as $\Im[G_{nn}(\omega)]=-\sum_m \Gamma_m|G_{mn}(\omega)|^2$. Then, 
the LDoS takes the form $\xi_1(\omega)=\frac{1}{\pi}\sum_m \Gamma_m|G_{m1}(\omega)|^2$, which allows us to further 
express it as a summation of the power densities per frequency unit $\Gamma_m|G_{m1}(\omega)|^2$ decaying into 
the environment with a rate $\Gamma_m$. Then, using $\Gamma_1=0$, Eq. (\ref{FGR}) can be written as 
\begin{equation}
\Gamma_{\text{FGR}}= 2v_e^2\sum_{m\neq 1}\Gamma_m|G_{m1}|^2 
\label{T_propto_xi_case}
\end{equation}
which can be further used to express the total emitted power as $P_{\text{total}}=P_{1,\rm{in}}=2v_e^2\sum_{m\neq 1}
\Gamma_m|G_{m1}|^2 |S_1^+|^2$. In the case of isolated resonant modes, the above expression predicts an increasing 
emissivity proportional to the $Q$-factor of the cavity modes. The latter is defined as $Q\equiv \frac{ \Re(
\omega_p)}{2\Im(\omega_p)}$ where $\omega_p$ is a pole of $G(\omega)$ and $2\Im(\omega_p)$ is the linewidth of the corresponding resonant mode.

Next, we quantify the emissivity enhancement associated with the presence of EPDs. To this end, we define the EPD-based
enhancement factor $\mathcal{F}(\kappa)$ as 
\begin{align}
    \mathcal{F}(\kappa)\equiv\frac{P_{\text{total}}[\omega_{\rm max}(\kappa)]/Q(\kappa)}{P_{\text{total}}[\omega_{\rm max}
    (\kappa_\infty)]/Q(\kappa_\infty)}
    = \frac{\xi[\omega_{\rm max}(\kappa)]/Q(\kappa)}{\xi[\omega_{\rm max}(\kappa_\infty)]/Q(\kappa_\infty)}
    \label{NPEFactor}
\end{align}
where $\omega_{\text{max}}(\kappa)$ indicates the frequency at which the total power is maximized for a given coupling $\kappa$. Above, $\kappa_\infty\gg \kappa_{\rm EPD}$ indicates the strong coupling domain for which the system demonstrates 
well-isolated resonances.

From the above, it is evident that the LDoS, and consequently the EPD-based enhancement factor, collapses 
to the analysis of the Green's function $G$. The latter, in the vicinity of an EPD-$N$, exhibits an anomalous behavior 
due to the coalescence of eigenmodes. Specifically,
\begin{align}
    G_{\rm EPD}(\omega)=\frac{\mathcal{I}}{\omega-\omega_{\rm EPD}}+...+\frac{M_N}{(\omega-\omega_{\rm EPD})^N} 
    \label{Green_func_EP}
\end{align}
where $M_k=(H-\omega_{\rm EPD}\mathcal{I})^{k-1}$ and $M_2$ is nilpotent of index $N$, thus guaranteeing the termination of the 
series~\cite{Heiss_2015,JW23}. Eq. (\ref{Green_func_EP}) indicates that $G(\omega)$ becomes highly sensitive to perturbations 
near an EPD-$N$. 
We can now write the LDoS as a series of Lorentzians of power $n\leq N$, i.e. 
\begin{equation}
    \xi_{\rm EPD}(\omega)=\frac{1}{\pi}\sum_{n=1}^Nc_n[L(\omega;\omega_{\rm EPD},\bar{\Gamma})]^n, 
\end{equation}
where $L(\omega;\omega_{\rm EPD},\bar{\Gamma})\equiv \frac{\bar{\Gamma}}{(\omega-\omega_{\rm EPD})^2 + \bar{\Gamma}^2}$ is a Lorentzian with linewidth described by $\bar{\Gamma}= \sum_{m\neq1}\Gamma_m/N$, the average of the decay rates of the resonators.
Higher-order terms lead to LDoS line shapes that are narrower and higher than the conventional Lorentzian LDoS, hence leading to 
enhanced emission.

\textit{Emissivity analysis for high-order EPDs --} We proceed with the comparison of the EPD-based enhancement factor 
for systems that support EPDs of order $N=2$ and $N=3$ (see Fig. \ref{fig1}). For the dimer $N=2$, the Hamiltonian $H_2$ 
supports an EPD-2 when $\kappa=\kappa_{\rm EPD}=\Gamma_2/2$. In this case, the two resonant modes acquire a degenerate value 
$\Re(\omega_p)=\omega_\pm=\omega_0$ with a linewidth $\Im(\omega_p)=\bar{\Gamma}=\Gamma_2/2$. The associated LDoS at resonator $n=1$ 
is a Lorentzian-squared $\xi_{\rm EPD}(\omega)=\Gamma_2/\pi \cdot [L(\omega;\omega_0, \bar{\Gamma})]^2$. In such a case, we obtain an EPD-based enhancement $\mathcal{F}_{\rm EPD-2}
\equiv \mathcal{F} (\kappa_{\rm EPD})\approx 4$ corresponding to the maximal possible enhancement for this configuration \cite{arkady,zin2,PZMHHRSJ17}, see 
Fig. \ref{fig1}(c).

Next, we analyze the emission for $N=3$, whose Hamiltonian $H_3$ has an EPD-3 at $\kappa_{1\rm ,EPD}=\sqrt{\bar{\Gamma}^3/\Gamma_3}$ and $\kappa_{2\rm ,EPD}=\sqrt{(2\Gamma_3-\Gamma_2)^3/(27\Gamma_3)}$, where $\bar{\Gamma}=(\Gamma_3+\Gamma_2)/3$ ~\cite{zin1}. Notice that 
for $\Gamma_2=2\Gamma_3$ the coupling $\kappa_{2\rm ,EPD}=0$, turning the trimer to an effective dimer with an EPD-2. 
Avoiding this condition results in the following general expression for the LDoS
\begin{equation}
    \xi_{\rm EPD}(\omega)=\frac{1}{\pi}\left(c_3 \cdot [L(\omega;\omega_0,\bar{\Gamma})]^3+c_2\cdot [L(\omega;\omega_0,\bar{\Gamma})]^2\right)\label{EP3_Lorentzian_cube}
\end{equation}
where $c_3=4[\Gamma_3\Gamma_2+\kappa_{2,\mathrm{EPD}}^2-2\bar{\Gamma}^2]$ and $c_2=[8\bar{\Gamma}^2-3(\Gamma_3
\Gamma_2+\kappa_{2,\mathrm{EPD}}^2)]/\bar{\Gamma}$. From Eq. (\ref{EP3_Lorentzian_cube}) we conclude 
that a dominant cubic Lorentzian term that will result in a greater emissivity enhancement (than the one associated 
with an EPD-2), requires that $c_3\gg c_2$. These two coefficients 
depend only on the dissipation rates $\Gamma_{2,3}$, which indicates the necessity of a judiciously designed 
dissipation profile of the cavity so that higher-order EPDs can further enhance the emissivity beyond the traditional Purcell prediction.

To evaluate the enhancement factor Eq. (\ref{NPEFactor}), we also need to evaluate the emissivity far away from the 
singularity, i.e., in the parameter domain $\kappa_{\infty}$ where the three resonances are isolated. This analysis requires 
identifying the perturbation scheme that controls the proximity to the EPD-3. We have chosen a specific coupling 
variation protocol such that 
\begin{equation}
\kappa_1=\sqrt{(\Gamma_2-\bar{\Gamma})\bar{\Gamma}+\frac{\bar{\Gamma}}{\Gamma_3-\bar{\Gamma}}\kappa_2^2};\quad 
\kappa_2=\kappa_{2\rm ,EPD}+\varepsilon
\label{kappa_SRE},
\end{equation}
where $\varepsilon$ is a perturbation parameter and $\Gamma_2<2\Gamma_3$ so that $\kappa_{2\rm ,EPD}>0$.  This perturbation 
scheme results in a detuning of the two ``lateral'' resonant modes $\omega_\pm$ away from the EPD-3 at $\omega_{\rm EPD}
=\omega_0$ that follows a square-root Newton-Puiseux series (the third ``central'' resonant mode remains at $\omega_0$) \cite{SM}. 
Importantly, using this perturbation protocol, we succeed in maintaining a constant $Q$-factor across the whole range of 
$\varepsilon$-values (see \cite{SM}).

We find that the maximum LDoS peak transitions from $\omega=\omega_0$ (central resonant mode) when $\Gamma_2<\Gamma_3$ to $\omega=
\omega_\pm(\varepsilon)$ (lateral modes) when $\Gamma_3<\Gamma_2<2\Gamma_3$. Hence, the EPD-based enhancement factor $\mathcal{F}_{\rm EPD-3}$ 
becomes
\begin{align}
 \mathcal{F}_{\rm EPD-3}=
 \begin{cases}
\frac{\Gamma_3}{\bar{\Gamma}^2(\Gamma_3-\bar{\Gamma})}\left(\frac{(2\Gamma_3-\Gamma_2)^3}{27\Gamma_3}+\Gamma_3\Gamma_2\right) & \text{if  } 
\Gamma_2\leq\Gamma_3 \\
\frac{2\Gamma_3}{\bar{\Gamma}^3}\left(\frac{(2\Gamma_3-\Gamma_2)^3}{27\Gamma_3}+\Gamma_3\Gamma_2\right) & \text{if } \Gamma_3< \Gamma_2<2\Gamma_3
        \end{cases}
   \label{F_SRE}
\end{align}
Eq. (\ref{F_SRE}) predicts that ${\cal F}_{\rm EPD-3}$ can be maximized for an appropriate choice of loss rates $\Gamma_3 = \Gamma_2$, 
where $\mathcal{F}^{\rm max}_{\rm EPD-3} = 7$. In this case, the three resonant emission peaks with equal height and width form a narrow LDoS 
profile at the EPD-3, see Fig. \ref{fig1}(d).

Finally, we point out that the CMT results are unaffected by the presence of small parasitic (radiative or dissipative) losses infesting 
the ``dark'' resonator.

\begin{figure}
    \includegraphics[width=0.47\textwidth]{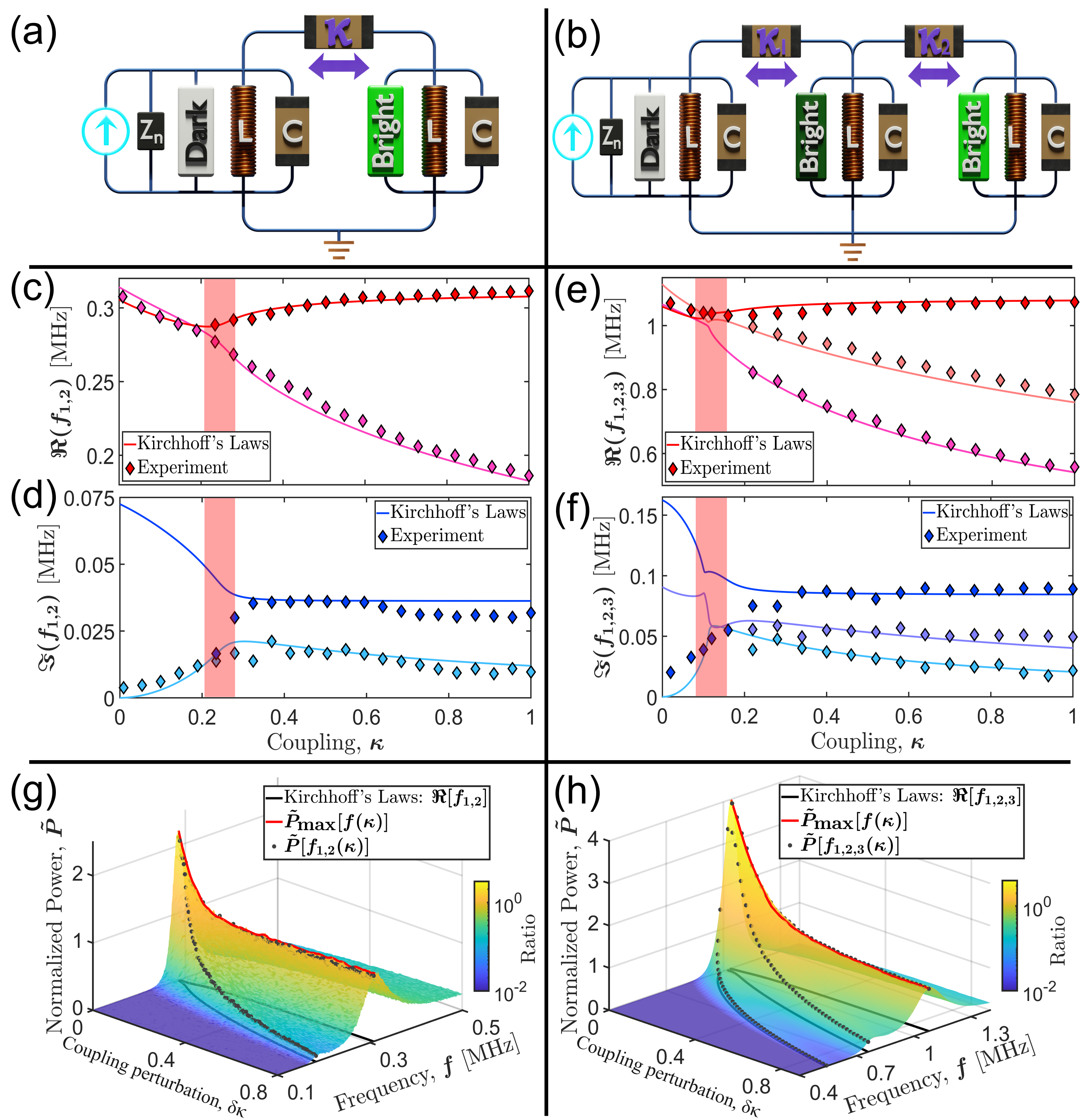}
    \caption{(a,b) Schematics of experimental RF cavities consisting of (a) two (dimer) and (b) three (trimer) RLC resonators with differential 
    loss. To emulate an emitter,  a monochromatic current source is placed in parallel with a load $Z_{n}=Z_{0}+1/(i\omega C_{e})$, which weakly couples it to the cavity via the dark resonator $n=1$. The emitted power is measured 
    at the transmission lines (TLs) attached to the other resonators. (c–f) Real and imaginary parts of eigenfrequencies for the dimer (c, d) and 
    trimer (e, f). Colored 
    diamonds: experimental data from best-fit of transmission spectra. Solid lines: eigenfrequencies of the RF cavities in the absence of coupling 
    to TLs, using Kirchhoff's equations. (g,h) Averaged experimental normalized emitted power versus frequency $f$ 
    and capacitive coupling detuning $\delta\kappa\equiv \kappa-\kappa_{ED}$ indicating $\tilde{P} \approx 2$ for the ED-2 and $\tilde{P} \approx 4$ for the ED-3. Solid black 
    lines on the bottom plane indicate the eigenfrequencies of the circuit extracted from Kirchhoff's equations, while black dots show the average 
    emitted power at such frequencies. The red line marks the power emitted from the most emissive supermode.
}
    \label{fig:circuits}
\end{figure}

\textit{Experimental RF platforms--} 
By utilizing an established mapping connecting CMT to electronic circuits \cite{H84,natcomm}, we employ our theoretical findings to guide the design of appropriate RF cavities.
We demonstrate our results using two RF configurations consisting of $N=2,3$ coupled RLC resonators 
(see Fig. \ref{fig:circuits}(a,b)). 
However, CMT predictions also have 
limitations. One such limitation is that for differential loss systems, CMT can exhibit exact EPDs, while RF cavities (described by Kirchhoff's 
laws) possess point-singularities only at the weak coupling limit between resonators. For RF scenarios where the inter-resonator 
coupling is moderate, the EPDs turn to exceptional domains (EDs) where the complex eigenfrequencies (and the associated eigenmodes) 
{\it nearly} coalesce. To further identify system parameters (inductances $L_{n},$ capacitances $C_{n},$ and resistances $R_{n}$) that lead 
to optimal EDs (specifically for the trimer $N=3$), we employed a Bayesian optimization  \cite{SSWAF16} that uses the CMT 
parameters as an initial guess. 

We found that an optimal ED for the dimer occurs for $L=240\text{ }\mu\text{H}$ and $C=1100 \text{ pF}$, corresponding 
to natural frequencies given by $f_{1,2}^{(0)}=f_{0}^{(0)}=\frac{1}{2\pi}\frac{1}{\sqrt{LC}}\approx0.308\text{ MHz}$. To implement 
the differential loss between the two RLC tanks, we have appropriately calibrated the setup so that resonator 2 is coupled to an additional 
parallel resistor $R_{2}=2\text{ k}\Omega$. For the trimer, the optimal parameters are $L=50\text{ }\mu\text{H}$ and $C_{1}=C_{3}=500\text{ pF}$ 
while $C_{2}=480\text{ pF}$. The corresponding natural frequencies are $f_{1,3}^{(0)}\approx1.01\text{ MHz}$ and $f_{2}^{(0)}\approx 
1.03\text{ MHz}$. The 
differential loss configuration was implemented by tuning the in-parallel resistances in each $RLC$ resonator to: $R_{1}=1 \text{ M}\Omega$ 
(minimal loss) using a fixed resistor, and using potentiometers to set $R_{2}=2\text{ k}\Omega$ and $R_{3}=1\text{ k}\Omega$. The coupling 
between the RLC tanks was maintained using fixed mica capacitors of various values $C_{\kappa}$, in parallel with each other, to produce 
coupling strengths $\kappa=C_{\kappa}/C_{i}\in (0,1]$ (mutual inductance coupling could be used as well). 

To emulate a radiative dipole emitting through the dark resonator of the RF cavity, we have {\it weakly} coupled a current source 
to each of our configurations. The latter has been experimentally implemented using the well-known equivalent generator theorem, 
which allows us to replace the circuit driven by a current generator with a circuit driven by a voltage source. Such a monochromatic 
source injects a $0.1$ mW signal into the ``dark'' resonator via a small capacitor $C_{e}=0.01\cdot C$ in series with a $Z_0=50\text{ }\Omega$ 
transmission line (TL). Finally, the emitted signal from the ``bright'' RLC resonators is collected via TLs coupled to a spectrum analyzer. 
 
\textit{Experimental measurements of resonant frequencies and linewidths -} Using the measured emitted power, we have extracted the 
associated resonance frequencies $f_n$ and their corresponding linewidths $\eta_n$ for various coupling 
strengths, see Fig. \ref{fig:circuits}(c-f). These correspond to the complex eigenfrequencies of the RF cavities. The ED is indicated with a red highlight, occurring at $\kappa_{\rm ED}\approx 0.23$ for 
$N=2$ (see Fig. \ref{fig:circuits}(c,d))) and at $\kappa_{\rm ED}\approx 0.113$ for $N=3$ (see Fig. \ref{fig:circuits}(e,f)). In both cases, 
the eigenfrequencies predicted by Kirchhoff's equations (solid lines) match with the experimental data (colored diamonds).

Our analysis of the parametric variation of the measured resonant frequencies $f_n$ with a coupling perturbation indicated that in 
both cases $N=2,3$, we have the same fractional (square) root expansion in the associated Newton-Puiseux series, despite the different underlying 
ED order (ED-2 vs. ED-3). We point out that for the trimer, the perturbation protocol takes the form of Eq. (\ref{kappa_SRE}), which 
simplifies to $\kappa_{1}=\kappa_{2}=\kappa_{\rm ED}+\delta\kappa $ under the specific differential loss configuration $\Gamma_2/\Gamma_3=1/2$ we have employed experimentally. 
In this case, Eq.~(\ref{F_SRE}) predicts $\mathcal{F}_{\rm EPD-3}=5$.

The experimental data indicate that $\eta_n$ remains approximately constant in the domain $\kappa > \kappa_{\rm ED}$; this is specifically 
true for the brightest supermode (upper curves in Fig. \ref{fig:circuits}(c,e)) that will be used to evaluate maximum emissivity. 
As a result, in both experimental configurations, the $Q$-factor of the brightest supermode, for which $f_{n}=f_{\rm ED}+\delta f$ where $\delta f\ll f_{\rm ED}$ also, remains approximately constant, i.e., $Q=(f_{\rm ED}+\delta f)/\eta\approx f_{\rm ED}/\eta$.

\textit{Power emission --} An overview of the measured relative emitted power ${\tilde P}(f;\kappa)={P (f;\kappa)\over P[f_{\rm max}
(\kappa_\infty)]}$ in the TLs for both $N=2 (3)$ is shown in Figs. \ref{fig:circuits}(g,h) as a density plot averaged over 
more than 10 different sets of measurements. The predicted resonant frequencies $f_n$ using Kirchoff's laws are shown as black 
lines in the $\delta\kappa-f$ plane. The grey dots mark the emission at these resonant frequencies, while the red line highlights the 
power emitted from the brightest supermode. The maximum measured values are ${\tilde P}(f;\kappa=\kappa_{\rm ED-N})\approx 2$ for 
$N=2$ and $\approx 4$ for $N=3$. Such results confirm that the order of the underlying ED affects the emissivity 
enhancement beyond the standard Purcell physics. Notice that, under the approximations that $Q(\kappa)=const.$ and Ohmic 
losses are small, ${\tilde P}(f;\kappa)$, is essentially the same as the non-Purcell EPD-based enhancement factor defined in Eq. (\ref{NPEFactor}). 
The latter has been estimated by CMT to be ${\mathcal F}=4$ for $N=2$ and ${\mathcal F}=5$ for $N=3$; which is consistent with 
the enhancement observed in our experiment.

\begin{figure}
    \centering
    \includegraphics[width=0.47\textwidth]{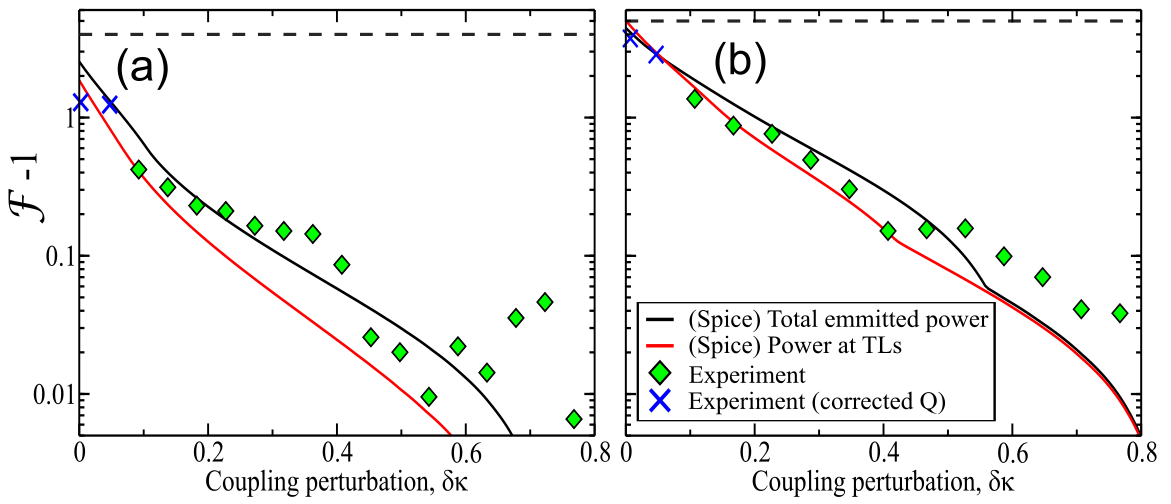}
    \caption{
EPD-based enhancement factor for (a) $N=2$ and $(b)$ $N=3$ configurations versus coupling perturbations $\delta\kappa$ 
away from the ED-$N$. The solid black (red) line corresponds to NGSpice simulations accounting for the total dissipated power 
(transmitted power at TLs), while green diamonds (blue crosses) represent the experimental power measurements rescaled 
with the experimental (simulated) $Q$-factor. In both configurations, the emitted power grows dramatically as $\delta \kappa 
\rightarrow 0$. The dashed horizontal lines indicate the EPD enhancement $\mathcal{F_{\text{CMT}}(\kappa_{\text{EPD}})}$ 
predicted by the CMT.}
    \label{fig:nonPurcell}
\end{figure}

To further scrutinize the influence of the two approximations on the EPD-based enhancement factor 
$\mathcal{F(\kappa)}$, we have performed NGSpice simulations that also take into account the dissipative channels due 
to the resistors and the small variations in the $Q$-factor. The dissipated power can be estimated from the transmitted power 
using a simple correction factor: $P_{R_n}=\frac{Z_0}{R_n}\left[ 1+ \frac{1} {(\omega C_e Z_0 )^2}\right] \cdot P_{\text{TL}_n}$ 
(see  \cite{SM}). In Fig. \ref{fig:nonPurcell} we report the NGSpice simulations for $\mathcal{F(\kappa)}$ 
that includes (excludes) the dissipative channels as a solid black (red) line. We also plot the experimental 
enhancement factor (green diamonds) evaluated using the emitted power at the TLs. The latter has been rescaled 
with the experimentally extracted $Q$-factor for moderate $\delta \kappa$. For $\delta \kappa
\approx 0$ the experimental evaluation of the $Q$-factor becomes noisy, and for this reason, we have used the NGSpice results. 
For the ED-2, the NGSpice simulations that include (exclude) the Ohmic dissipated power yield an enhancement of 
$\mathcal{F_{\text{dimer}}^{(\text{total})}}\approx 4$ ($\mathcal{F_{\text{dimer}}^{(\text{TL})}}\approx 3$). We also
find that an ED-3 significantly boosts the emissivity enhancement, resulting in $\mathcal{F_{\text{trimer}}^{
(\text{total})}}\approx \mathcal{F_{\text{trimer}}^{(\text{TL})}}\approx 5$. These results are consistent with the 
experimental findings, which give $\mathcal{F_{\text{ED-2}}^{(\text{TL})}}\approx 2.3$ and $\mathcal{F_{\text{EPD-3}}^{(\text{TL})}}\approx 4.8$.

\textit{Conclusions -} We have presented an experimental demonstration of emissivity enhancement beyond the typical Purcell factor, which is proportional to the $Q$-factor, near higher-order 
$(N=3)$ EDs in an RF coupled RLC-resonator system. Our experimental findings agree with theoretical predictions that 
indicate that such power emissivity enhancement is a consequence of a judiciously designed spatial dissipation profile that promotes 
a cubic power Lorentzian emission line shape as opposed to a square power Lorentzian term. These results could be used to improve the 
design of high-emission radio-antennas that currently rely on standard non-degenerate resonances. The emission properties 
associated with these higher-order degeneracies can be exploited in fields other than RF, ranging from high-power low-coherence 
light sources, and sources with tunable coherence, (quantum) sensing and imaging.

\begin{acknowledgments}
L.F.A. and T.K. acknowledge partial support from MPS Simons Collaboration via grant No. SFI-MPS-EWP-00008530-08. N.W., A.S., and T.K. acknowledge partial support from the Department of Energy grant No. DE-SC0024223, Office of Naval Research MURI grant No N000142412548, and from NSF/RINGS under Grant No. ECCS-2148318. L.F.A. acknowledges partial support from CONICET and (ex)MINCyT Grant No. CONVE-2023-10189190-FFFLASH.
\end{acknowledgments}

\end{document}